\documentclass[onecolumn,showpacs,preprintnumbers,amsmath,amssymb]{revtex4}

\usepackage{graphicx}% Include figure files
\usepackage{dcolumn}% Align table columns on decimal point
\usepackage{bm}% bold math

\begin{document}

\title{Exact time-averaged thermal conductance for small systems:
Comparison between direct calculation and Green-Kubo formalism}

\author{W. A. M. Morgado}
 \email{welles@fis.puc-rio.br}
\affiliation{Departamento de F\'{i}sica, Pontif\'{i}cia
Universidade Cat\'{o}lica\\ and National Institute of Science and Technology for Complex Systems\\ 22452-970, Rio de Janeiro, Brazil}%

\author{D. O. Soares-Pinto} \email{dosp@cbpf.br}
\affiliation{Centro Brasileiro de Pesquisas
F\'{\i}sicas, Rua Dr.
Xavier Sigaud 150, 22290-180, Rio de Janeiro, Brazil}%
%\date{\today}

\begin{abstract}
In this paper, we study exactly the thermal conductance for a low
dimensional system represented by two coupled massive Brownian
particles, both directly and  via a Green-Kubo expression. Both
approaches give exactly the same result. We also obtain exactly the
steady state probability distribution for that system by means of
time-averaging.
\end{abstract}


\maketitle

\section{Introduction}

Exact results, in physics, play an important and useful role as a
reference for other methods. For instance, they can be used to study
specific features of models which are not easily accessible to
approximative methods, such as computer simulations. However,
non-trivial exact results are few and difficult to come by: in the
literature of  transport phenomena there are not many exact
calculations for transport coefficients based on the mechanical
parameters of the systems under observation
\cite{1963_PhysRev_131_964,1967_JMP_8_1073,1971_JMathPhys_12_1686,
1970_PRA_4_1086,1975_PRB_6_2164,1990_PRA_42_3278,2003_PhysRep_377_1,2004_JSP_116_783,2005_PRL_95_094302}.
Indeed, the calculation of transport coefficients is one of the most
important goals of non-equilibrium physics.

In fact, the rigorous derivation of Fourier's Law for bulk
Hamiltonian systems is still under debate~\cite{2004_JSP_116_783}.
The soluble harmonic models used to evaluate the thermal conductance
do not reproduce the necessary scattering of the energy unless local
thermal reservoirs (which act in part as effective scatterers for
the energy incoming on them) are coupled to the bulk sites. Other
models, such as mass disordered ones have been proposed but there is
evidence that mass disorder alone will not give rise to Fourier's
Law in 2D~\cite{2005_PRL_95_094302}. The presence of anharmonicity
on the coupling level would probably be sufficient for energy to
scatter and diffuse from site to neighboring site. However, the
technical details for obtaining exact rigorous results are, as far
as we know, too difficult to overcome at the present
level~\cite{2004_JSP_116_783}.

On the other hand, for Equilibrium Statistical Mechanics the
probability distribution for a given system can be found, given that
it obeys Liouville's theorem~\cite{livro_huang} and the external
macroscopic constraints~\cite{livro_reif}, by means of an ensemble
of points in phase space, when that system is ergodic. However,
ergodicity is not a necessary condition for obtaining the long times
the stationary probability distribution  since it is always
possible, at least in principle, to obtain the time-average for any
physical quantity during the realization of an actual experiment.

What distinguishes time averaging from other exact methods derived
from the solution of a Fokker-Planck formalism is that time
averaging can take into account, exactly, all the orders of the
moments of the dynamical variables. The Fokker-Planck formalism is
exact only to the second order moments while time average is akin to
the stationary solution of the Kramers-Moyal equation correct at all
orders of moments~\cite{livro_vankampen}. For simplicity sake we use
Gaussian white noise in the present work but the method is readily
generalizable to any type of noise, given that all its moments are
known.

Our present goal is to study exactly the validity, and consistency,
of some methods used in the derivation of transport coefficients
(namely the thermal conductance) for small classical systems. The
reasons for using small systems in our model are manifold. Firstly,
exact calculations become feasible. Secondly, macroscopic transport
properties associated with large systems must have a mechanical
counterpart in small systems. Macroscopic flows of mass, momentum or
energy are the effect of the averaging of the action of microscopic
forces, and work.  Thirdly, small systems are interesting {\it per
se}. There are difficulties inherent to small systems due to the
fact that one cannot take the thermodynamic limit that averages out
many problems associated with solving the dynamics of large systems,
similarly to the Law of Large Numbers that arises from the summation
of many random variables~\cite{livro_feller}.  We can also take into
account exactly the effects of the inertia and of a possible
non-Markovian nature for the noise.

For deriving  the exact time-averaged thermal conductance, we
choose a simple system which is capable of non-trivially
transmitting  heat between its constitutive parts, via mechanical
work, when  subjected to a gradient of (white noise) temperature:
a system of two coupled Brownian particles (BP) with the coupling
constant $k$ acting as the sole information channel between the
particles. In the present case, the system dynamics is linear and
the linear response treatment  shall be proven to be exact, as we
will see in the following.

We can apply a method previously used by the
authors~\cite{2006_PhysA_365_289,2008_PRE_77_011103} to the study of
thermal conduction between the coupled BP system.  The exact
time-averaging method of
References~\cite{2006_PhysA_365_289,2008_PRE_77_011103} is capable
of obtaining  exactly the stationary probability distribution for
Brownian particles submitted to white and colored noise. In
particular, by submitting a single massive Brownian particle (BP) to
two different thermal contacts, at distinct temperatures (similarly
to some glasses that are subjected to thermal vibrations and
structural modifications represented by distinct noise functions at
different time-scales~\cite{2004_RMP_76_785}), we can keep it from
reaching thermal equilibrium~\cite{2008_PRE_77_011103}.

Since the present model is effectively zero dimensional, the thermal
conductance between the Brownian particles is defined simply as the
energy flow per unit time per temperature difference between the
particles,i.e., the conductive flow of energy  (for particle 1, from
particle 2) is defined as $j_{1,2}= -\kappa\,(T_{1,2}-T_{2,1}), $
where $\kappa$ is the inter-particle thermal conductance in
first-order approximation. Indeed, there has been some recent
developments in treating finite systems that can be adapted to the
problem under study~\cite{2008_AdvPhys_57_457,2009_JSTAT_L03001}. In
that case, the transport coefficient $\kappa$ is obtained via a
convenient Green-Kubo formulation. The calculation of the transport
coefficients by this method can be an interesting starting point for
the study of more complex models, such as polymers subjected to
gradients of
temperature~\cite{2004_JSP_116_783,1967_JMP_8_1073,1970_PRA_4_1086}.
It will also provide an important test for the choice of flow
variable appropriate for such models. However, in order to avoid the
rather artificial construction of an ensemble of reservoirs that
need to be coupled to the particles along the linear polymer
(harmonic crystal), a generalization of the method will be needed to
include non-linearities on the potential.  This way, a much more
realistic picture of thermal conduction will be obtained.

The thermal conductance between two particles is not a well defined
macroscopic quantity since we are far from the thermodynamic limit
and cannot define a macroscopic (and diffusive) flow of heat.
However, it is clear that if a (classic) macroscopic system is
partitioned  into two parts, energy conduction is realized by the
interactions (work) at the interface.

Furthermore, we add a periodic variation of the temperature of the
Brownian particles. This is an interesting effect that can lead to
the appearance of currents for systems presenting asymmetries in the
potential energy~\cite{1997_PhysD_109_17,1996_PLA_215_26}.  Periodic
oscillations of different types are capable of creating currents
~\cite{1990_PRA_6_2977,1994_PRE_5_3626} in the case of zero average
forces acting on the particles. The combination of ratchet-type
potential energy and periodic time oscillation for the temperature
has been extensively studied
\cite{2008_EPL_84_40009,1996_PLA_215_26} .

This paper is organized as follows: In Section II we define the
model. In Section III we explain the method of time-averaging and
show the main contributions to the probability distribution. In
Section IV we calculate the time-averaged  steady-state
distribution for the non-equilibrium conditions. In Section V we
obtain the thermal conductance and in Section VI we discuss our
main conclusions.

\section{Exactly solvable model}

Our model consists of two massive Brownian particles (BP) coupled by
an harmonic potential and subjected to white noise at distinct
temperatures. This could be interpreted as two atoms in a crystal,
coupled by a harmonic potential.

Despite the reduced number of variables, the present system contains
the main ingredients of more complex models. In it, we can define
the energy transfer as the microscopical work, that in macroscopical
systems become the internally transferred heat. In the following, we
describe the model in detail and, using time-average
techniques~\cite{2006_PhysA_365_289,2008_PRE_77_011103}, we
calculate exactly the probability distribution for the relevant
Brownian variables.

\subsection{Langevin-type equation}
The system composed by two coupled punctual and massive BPs is
described by the equations:
\begin{eqnarray}
\dot x_{\alpha}(t)&=&v_{\alpha}(t), \label{eq.03}\\
m_{\alpha}\,\dot v_{\alpha}(t)&=&
-k\,(x_{\alpha}(t)-x_{\beta}(t))-k^{\prime}\,x_{\alpha}(t) -
\gamma_{\alpha}v_{\alpha}(t)
+ \eta_{\alpha}(t). \label{eq.04}%\\
\end{eqnarray}

Gaussian behavior is to be expected for the probability
distribution for the time-averaged stationary state, according to
previous published works (in special see sections 1.3.E.2 and
2.2.E.2 in Ref.~\cite{1982_PhysRep_88_207}).

For simplicity, we make: $m_{1}= m_{2}= m$, $k^{\prime\prime} =
k^{\prime}$, and $\gamma_{1} = \gamma_{2} = \gamma$. Thus the
equations can be written as:
\begin{equation}\label{eq.07}
m\,\ddot x_{\alpha}(t)=
-k\,(x_{\alpha}(t)-x_{\beta}(t))-k^{\prime}\,x_{\alpha}(t) -
\gamma\dot x_{\alpha}(t) + \eta_{\alpha}(t),
\end{equation}
%and
%\begin{equation}\label{eq.08}
%m\,\ddot x_{2}(t)= -k\,(x_{2}(t)-x_{1}(t))-k^{\prime}\,x_{2}(t)-
%\gamma\dot x_{2}(t) + \eta_{2}(t),
%\end{equation}
where $\alpha, \beta = 1, 2$, $\alpha \neq \beta$, and the initial
conditions are:
\[
x_1(0) = x_2(0) = v_1(0) = v_2(0) = 0.
\]

\subsection{Noise properties}

Both white Gaussian noise terms can be defined in terms of their two
lowest two cumulants:
\begin{eqnarray}
\langle\eta_{\alpha}(t)\rangle &=& 0, \label{eq.09}\\
\langle\eta_{\alpha}(t)\eta_{\beta}(t^{'})\rangle &=&
2\,\gamma\,T_{\alpha}(t)\,\delta_{\alpha\beta}\,\delta(t-t'),
\label{eq.11}
\end{eqnarray}
where the modulated temperatures above are given by
\begin{equation}\label{eq.12}
T_{\alpha}(t) = \bar{T}_{\alpha}\,\left[1 +
A_{\alpha}\,\sin(\omega_{\alpha}\,t)\right]^{2},
\end{equation}
for $\alpha = 1,2$ and $|A_{\alpha}|<1$.

The oscillating temperatures, in other models,  can induce very
interesting effects such as sending heat fluxes against gradients
of temperature~\cite{2008_EPL_84_40009} or directed fluxes of
particles in periodic potentials~\cite{1997_PhysD_109_17,
1996_PLA_215_26, 1996_LectNotesPhys_476_294, 2002_AppPhysA_75_169,
2009_RMP_81_387}.

%
%     C
%
\subsection{Laplace transformations}

Taking the Laplace transformations of Eqs.(\ref{eq.03}) and
(\ref{eq.04}) yields

\begin{eqnarray}
(m\,s^{2}+\gamma\,s+k+k^{\prime})\,\tilde{x}_{1}(s) &=&
k\,\tilde{x}_{2}(s)+ \tilde{\eta}_{1}(s), \label{eq.13}\\
\tilde{v}_{1}(s) &=& s\,\tilde{x}_{1}(s), \label{eq.14}\\
(m\,s^{2}+\gamma\,s+k+k^{\prime})\,\tilde{x}_{2}(s) &=&
k\,\tilde{x}_{1}(s)+ \tilde{\eta}_{2}(s), \label{eq.15}\\
\tilde{v}_{2}(s) &=& s\,\tilde{x}_{2}(s). \label{eq.16}
\end{eqnarray}

Defining $\Gamma(s) \equiv m\,s^{2}+\gamma\,s+k+k^{\prime}$ and
rearranging Eqs.(\ref{eq.13}) to (\ref{eq.16}), one finds that:
\begin{eqnarray}
\tilde{x}_{1}(s) &=& \Lambda(s)\,\tilde{\eta}_{1}(s)+
\Delta(s)\,\tilde{\eta}_{2}(s), \label{eq.17}\\
\tilde{v}_{1}(s)&=& s\,\Lambda(s)\,\tilde{\eta}_{1}(s)+
s\,\Delta(s)\,\tilde{\eta}_{2}(s), \label{eq.18}\\
\tilde{x}_{2}(s) &=& \Delta(s)\,\tilde{\eta}_{1}(s)+
\Lambda(s)\,\tilde{\eta}_{2}(s), \label{eq.19}\\
\tilde{v}_{2}(s) &=& s\,\Delta(s)\,\tilde{\eta}_{1}(s)+
s\,\Lambda(s)\,\tilde{\eta}_{2}(s), \label{eq.20}
\end{eqnarray}
where:
\begin{eqnarray}
\Lambda(s) &\equiv & \frac{\Gamma(s)}{\Gamma^{2}(s)-k^{2}}, \label{eq.21}\\
\Delta(s) &\equiv & \frac{k}{\Gamma^{2}(s)-k^{2}}. \label{eq.22}
\end{eqnarray}

\begin{figure}[tbh]
\begin{center}
\includegraphics[width=0.5\columnwidth,angle=0]{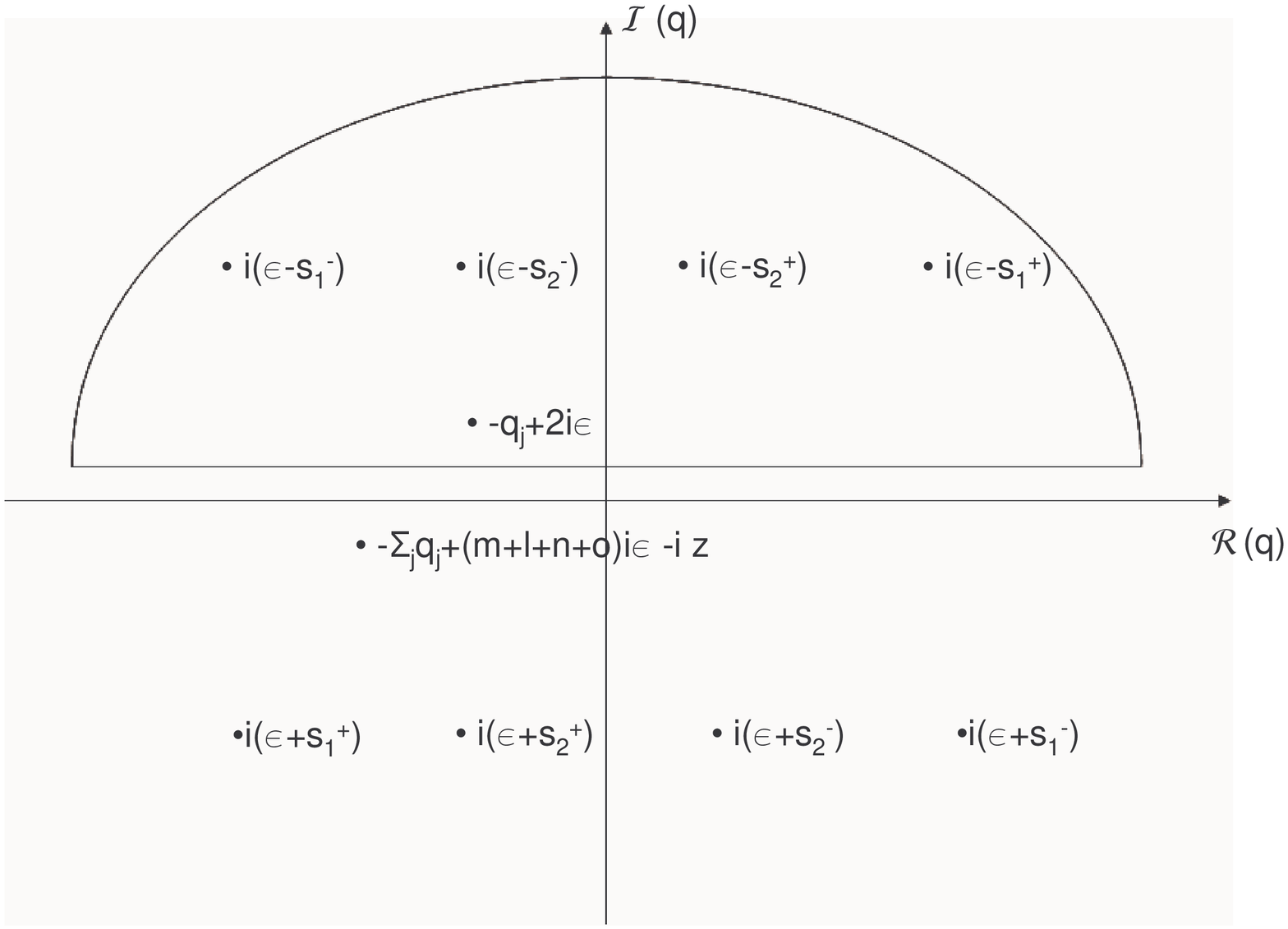}
\end{center}
\caption{Integration path for the equilibrium distribution,
Eq.(\ref{eq.25}).}\label{fig1}
\end{figure}

The Laplace transformation for the independent noise variables is
given by ($\alpha=1,2$):
\begin{eqnarray}
\frac{\langle\tilde{\eta}_{\alpha}(iq_{i}+\epsilon)\,
\tilde{\eta}_{\alpha}(iq_{j}+\epsilon)\rangle}{2\,\gamma\,\bar{T}_{\alpha}}
&=&\left[\frac{1}{i(q_i+q_j)+ 2\epsilon}
+\frac{2\,A_{\alpha}\omega_{\alpha}}{[i(q_i+q_j)+
2\epsilon]^{2}+\omega_{\alpha}^{2}}\right. \nonumber\\
&+&\left.\frac{2\,A_{\alpha}^{2}\omega_{\alpha}^{2}}{[i(q_i+q_j)+
2\epsilon]\left([i(q_i+q_j)+2\epsilon]^{2}+
4\,\omega_{\alpha}^{2}\right)}\right]\label{eq.23}
\end{eqnarray}
All integration paths are the same and shown in Fig.\ref{fig1}.

%%%%%%%%%%%%%%%%%%%%%%%%%%%%%%%%%%%%%%%%%%%%%%%%%%%
%%%%%%%%%%%%%%%%%%%%%%%%%%%%%%%%%%%%%%%%%%%%%%%%%%%

\section{Probability distribution}

Here we show some of the steps (more detail can be found in
references~\cite{2006_PhysA_365_289,2008_PRE_77_011103}) to obtain
the expression for the instantaneous probability distribution for
the system of coupled Brownian particles.

Time averaging, for a supposedly convergent distribution, is defined
and calculated as in Refs.
\cite{2006_PhysA_365_289,2008_PRE_77_011103}:
\begin{eqnarray*}
\bar{f}& =& \lim_{\Omega\rightarrow\infty}
\frac{1}{\Omega}\int_{0}^{\Omega}dt\,f(t) = \lim_{z\rightarrow 0^{+}} z\int_{0}^{\infty}dt\,e^{-zt}\,f(t)
\end{eqnarray*}

From the definition for the instantaneous probability distribution:
\begin{eqnarray}
p(x_1,v_1,x_2,v_2,t) &=&
\langle\delta(x_1-x_1(t))\,\delta(v_1-v_1(t))\,
\delta(x_2-x_2(t))\,\delta(v_2-v_2(t)) \rangle,\label{eq.24}
\end{eqnarray}
it is possible to show that
\cite{2006_PhysA_365_289,2008_PRE_77_011103}:
\begin{eqnarray}
p^{ss}(x_1,v_1,x_2,v_2) &=&
\lim_{z\rightarrow0}\lim_{\epsilon\rightarrow0}
\int_{-\infty}^{+\infty} \frac{dQ_1}{2\pi}\frac{dQ_2}{2\pi}
\frac{dP_1}{2\pi}\frac{dP_2}{2\pi}
e^{iQ_1x_1+iP_1v_1+iQ_2x_2+iP_2v_2}\times\nonumber\\&\times&
\sum_{l,m,n,o=0}^{\infty} \frac{(-iQ_1)^{l}}{l!}
\frac{(-iP_1)^{m}}{m!}\frac{(-iQ_2)^{n}}{n!}
\frac{(-iP_2)^{o}}{o!}\nonumber\times\\
&\times & \int_{-\infty}^{+\infty} \prod_{f=1}^{l}
\frac{dq_{1f}}{2\pi}\prod_{h=1}^{m}
\frac{dp_{1h}}{2\pi}\int_{-\infty}^{+\infty} \prod_{j=1}^{n}
\frac{dq_{2j}}{2\pi}\prod_{k=1}^{o} \frac{dp_{2k}}{2\pi}
\times \nonumber\\
& &\hspace{-2.0 cm}\times\,\,\,
\frac{z^{m+o+1}}{z-\left[\sum_{a=1}^{l}(iq_{1a}+\epsilon)
+\sum_{b=1}^{m}(ip_{1b}+\epsilon)+\sum_{c=1}^{n}(iq_{2c}+\epsilon)
+\sum_{d=1}^{o}(ip_{2d}+\epsilon)\right]}
\times\nonumber\\
&\times &  \langle \prod_{f=1}^{l} \tilde{x}_1(iq_{1f}+\epsilon)
\prod_{h=1}^{m}\tilde{v}_1(ip_{1h}+\epsilon) \prod_{j=1}^{n}
\tilde{x}_2(iq_{2j}+\epsilon) \prod_{k=1}^{o}\tilde{v}_2
(iq_{2k}+\epsilon)\rangle, \label{eq.25}
\end{eqnarray}
where the integration paths for the $(q,p)$ variables are given in
Fig.\ref{fig1}.

An interesting case is the study of the average $\langle
x_1^2(t)\rangle$. The time periodicity of the noise
[$T_{1,2}(t+\mathcal{T})=T_{1,2}(t)$] gets translated into a
periodicity of distribution [$p(x,t+\mathcal{T})=p(x,t)$], and by
consequence also of the averages for the variables, e.g. $\langle
x_1^2(t+\mathcal{T})\rangle = \langle x_1^2(t)\rangle$. This is
consistent with a Fokker-Planck treatment of the distribution
found in the literature~\cite{1997_PhysD_109_17, 1996_PLA_215_26,
1996_LectNotesPhys_476_294, 2002_AppPhysA_75_169,
2009_RMP_81_387}. We now show that the instantaneous distribution
described by Eq.(\ref{eq.25}) is indeed consistent with
periodicity in time.

From reference~\cite{1996_PLA_215_26}, it is clear that $\langle
x_1^2(t)\rangle$ is periodic in time with period $\mathcal{T}$:
\[
\langle x_1^2(t+\mathcal{T})\rangle
=\int\,dx_1\,p(x_1,t+\mathcal{T})\,x_1^2
=\int\,dx_1\,p(x_1,t)\,x_1^2 = \langle x_1^2(t)\rangle,
\]
where the periodicity of the probability distribution directly
implies that of the average. We assume that $T_2=0, k=0$ so the
present model and the one from~\cite{1996_PLA_215_26} coincide.

The average (for long time, after the memory of the initial
conditions has already faded out) reads:
\begin{eqnarray*}
\langle x_1^2(t)\rangle &=& \int\,dx_1\,p(x_1,t)\,x_1^2 \\
&=& \lim_{\epsilon\rightarrow0} \int_{-\infty}^{+\infty}
\frac{dq_{1}}{2\pi}\frac{dq_{2}}{2\pi}
e^{t\left[iq_{1}+iq_{2}+2\epsilon\right]} \langle
\tilde{x}_1(iq_{1}+\epsilon) \tilde{x}_1(iq_{2}+\epsilon) \rangle
\end{eqnarray*}

By integrating the last equation above over the poles of
$\langle\tilde\eta_1\tilde\eta_2\rangle$ we observe that apart form
the term of order $O(A^0)$, the terms carrying the contribution from
the sine bring a dependence such that
$iq_{1}+iq_{2}+2\epsilon=\pm\,i\, \omega$. By obtaining the residue,
we are left with exponential terms of the form
$e^{\pm\,i\,\omega\,t},$ which are periodic in time with period
$\mathcal{T}$.

An interesting aspect of periodically varying noise is that it
modulates the time behavior of distributions and averages, i.e.,
there are no stationary constant values for any of the moments of
the Brownian variables: the moments are  periodic functions of time.
In this case, the time average we use corresponds to averages of
these quantities taken over a period of the noise for very long
observation times. In general lines, the next steps consist into
expressing the Laplace transforms for the dynamical variables
$(x,v)$ into functions of the averages of the Laplace transforms of
the noise.

%
%       IV
%
\section{Time-averaged steady state distribution}

\subsection{Contributing terms}

The main contributions for the probability shown in
Eq.(\ref{eq.25}) comes from a typical integration is of the form
\cite{2006_PhysA_365_289, 2008_PRE_77_011103}:
\begin{equation}\label{eq.26}
\mathcal{C}_{\tilde{\alpha}_{r}\,\tilde{\beta}_{s}} =
\int_{0}^{\infty}\frac{dq_{i}}{2\,\pi}\frac{dq_{j}}{2\,
\pi}\frac{z}{z-i(q_{i}+q_{j}+2\epsilon+\oslash)}
\langle\tilde{\alpha}_{r}(iq_{i}+\epsilon)\,\tilde{\beta}_{s}(iq_{j}+
\epsilon)\rangle,
\end{equation}
where $\oslash = \left[\sum_{a=1,a\neq\,i}^{l}(iq_{a}+\epsilon)
+\sum_{b=1,b\neq\,j}^{n}(iq_{b}+\epsilon) \right]$,
$\alpha,\beta=x,v$, and $r,s=1,2$.

To understand what causes a term in Eq.(\ref{eq.25}) to contribute
to the time-averaged steady state probability distribution it is
necessary to observe that for a typical integration, such as the
above term, there is a factor $I(z)$
\[
I(z) = \frac{z}{z-i(q_{i}+q_{j}+2\epsilon+\oslash)},
\]
that, in the limit $z\rightarrow0$, will vanish if there is the
presence of any finite terms on its denominator  (due to the residue
calculations around the poles of the rest of the integrand). Only
integrations that eliminate all the pairs of q's in the denominator
of $I(z)$ will transform it into $I(z) = z/z=1$. This is a necessary
condition for any of the integrations done below to contribute to
the probability distribution. Each integration brings the
corresponding multiplicative factors that need to be dealt with.

We define:
\begin{eqnarray}
\mathcal{Q}_{\Lambda\,\Lambda} &=& \frac{k+k^{\prime}}
{4\,\gamma\,(k^{\prime\,2}+2\,k\,k^{\prime})}+
\frac{\gamma}{4\,[m\,k^{2}+(k+k^{\prime})\,\gamma^{2}]},\label{eq.27}\\%\label{QGG}\\
\mathcal{Q}_{\Delta\,\Delta} &=& \frac{k^{2}\,
[m(k+k^{\prime})+\gamma^{2}]}{4\,\gamma\,k^{\prime}\,
(2\,k+k^{\prime})[m\,k^{2}+(k+k^{\prime})\,\gamma^{2}]},\label{eq.28}\\%\label{Qkk}\\
\mathcal{Q}_{\Delta\,\Lambda} &=& \mathcal{Q}_{\Lambda\,\Delta} \,
= \,
\frac{k}{4\,\gamma\,k^{\prime}\,(2\,k+k^{\prime})},\label{eq.29}\\%\label{QkG}\\
\mathcal{R}_{\Lambda\,\Lambda} &=& \frac{1}{4}
\left\{\frac{m\,k^{2}+2\,(k+k^{\prime})\,
\gamma^{2}}{m\,k^{2}+(k+k^{\prime})\,\gamma^{2}}\right\},\label{eq.30}\\%\label{RGG}\\
\mathcal{R}_{\Delta\,\Delta} &=& \frac{1}{4}
\left\{\frac{m\,k^{2}}{m\,k^{2}+
(k+k^{\prime})\,\gamma^{2}}\right\},\label{eq.31}\\%\label{Rkk}\\
\mathcal{H}  & = &\frac{k\,\gamma\, }
{4\,\left[m\,k^{2}+\gamma^{2}(k+k^{\prime})\right]}.\label{H}
\end{eqnarray}

Thus,  the contributing terms will be:
\begin{eqnarray}
\mathcal{C}_{\tilde{x}_{1}\,\tilde{x}_{1}} &=&
\frac{z}{z-i\,\oslash}\,
\left\{\gamma\,\bar{T}_{1}\,(2+A_{1}^{2})\mathcal{Q}_{\Lambda\,\Lambda}
+ \gamma\,\bar{T}_{2}\,
(2+A_{2}^{2})\mathcal{Q}_{\Delta\,\Delta},\right\},\label{eq.32}\\
\mathcal{C}_{\tilde{x}_{2}\,\tilde{x}_{2}} &=&
\frac{z}{z-i\,\oslash}\,\left\{\gamma\,
\bar{T}_{1}\,(2+A_{1}^{2})\mathcal{Q}_{\Delta\,\Delta} +
\gamma\,\bar{T}_{2}\,
(2+A_{2}^{2})\mathcal{Q}_{\Lambda\,\Lambda},\right\},\label{eq.33}\\
\mathcal{C}_{\tilde{x}_{1}\,\tilde{x}_{2}} &=&
\frac{z}{z-i\,\oslash} \left\{\gamma\,\bar{T}_{1}(2+A_{1}^{2}) +
\gamma\,\bar{T}_{2}
(2+A_{2}^{2})\right\} \mathcal{Q}_{\Delta\,\Lambda},\label{eq.34}\\
\mathcal{C}_{\tilde{v}_{1}\,\tilde{v}_{1}} &=&
\frac{z}{z-i\,\oslash}\, \left\{\frac{\bar{T}_{1}}{m}
\,(2+A_{1}^{2})\mathcal{R}_{\Lambda\Lambda}
+\frac{\bar{T}_{2}}{m}\,(2+A_{2}^{2})\mathcal{R}_{\Delta\Delta}\right\},\label{eq.35}\\
\mathcal{C}_{\tilde{v}_{2}\,\tilde{v}_{2}} &=&
\frac{z}{z-i\,\oslash}\,\left\{\frac{\bar{T}_{1}}{m}
\,(2+A_{1}^{2})\mathcal{R}_{\Delta\Delta}
+\frac{\bar{T}_{2}}{m}\,(2+A_{2}^{2})
\mathcal{R}_{\Lambda\Lambda}\right\}.\label{eq.36}\\
\mathcal{C}_{\tilde{x}_{1}\,\tilde{v}_{2}} &=&
\frac{z}{z-i\,\oslash}\,\mathcal{H}\left[\bar{T}_{1}(2+A_{1}^{2})
-\bar{T}_{2}(2+A_{2}^{2})\right]\\
\mathcal{C}_{\tilde{x}_{2}\,\tilde{v}_{1}} &=&
\frac{z}{z-i\,\oslash}\,\mathcal{H}\left[\bar{T}_{2}(2+A_{2}^{2})
-\bar{T}_{1}(2+A_{1}^{2})\right]
\end{eqnarray}

The other possible terms vanish:
\begin{equation}\label{eq.37}
\mathcal{C}_{\tilde{v}_{1}\,\tilde{v}_{2}} =
\mathcal{C}_{\tilde{x}_{1}\,\tilde{v}_{1}} =
\mathcal{C}_{\tilde{x}_{2}\,\tilde{v}_{2}} = 0,
\end{equation}
since they are integrations of products of  odd functions of $q$'s with even functions of  $q$'s.

Notice that the effect of modulation is to scale the temperature
value by a factor of $1+A^2/2$. This is exactly what is obtained
by taking the time average of $T(1+A\sin(\omega\,t))^2$:
\[
\overline{T(1+A\sin(\omega\,t))^2}=T(1+A^2/2).
\]

We notice that the coupling term $k$ is responsible for the non-zero
values of $\mathcal{Q}_{\Lambda\Delta}$,
$\mathcal{Q}_{\Delta\Delta}$, $\mathcal{R}_{\Delta\Delta}$ and
$\mathcal{H}$.

\subsection{Exact solution for the time-averaged steady state distribution}

Using the results obtained above, it is possible to obtain that
Eq.(\ref{eq.25}) is:

\begin{equation}\label{PSSW}
p^{ss}(x_1,v_1,x_2,v_2) =
\int_{-\infty}^{+\infty} \frac{dQ_1}{2\pi}\frac{dQ_2}{2\pi}
\frac{dP_1}{2\pi}\frac{dP_2}{2\pi}
e^{iQ_1x_1+iP_1v_1+iQ_2x_2+iP_2v_2}\,\mathcal{W}(Q_1, P_1, Q_2, P_2)
\end{equation}
where:

\begin{eqnarray}
\mathcal{W}(Q_1, P_1, Q_2, P_2) &=& \sum_{M=0}^{\infty}
\sum_{N=0}^{\infty} \sum_{S=0}^{\infty} \sum_{T=0}^{\infty}
\frac{(iQ_1)^M}{M!} \frac{(iQ_2)^N}{N!} \frac{(iP_1)^S}{S!}
\frac{(iP_2)^T}{T!} \left<\tilde{x}_1^M \tilde{x}_2^N\tilde{v}_1^S
\tilde{v}_2^T\right>\nonumber\\
&=&\exp\left\{-\mathcal{H}\left[\bar{T}_{1}(2+A_{1}^{2}) -
\bar{T}_{2}(2+A_{2}^{2})\right]
\left[Q_1P_2-Q_2P_1\right]\right\}\times \nonumber
\\&\times&\exp\left\{-\left(\frac{Q_1^2\mathcal{Q}_{\Lambda\,\Lambda}+
2Q_1Q_2\,\mathcal{Q}_{\Delta\,\Lambda}
+Q_2^2\mathcal{Q}_{\Delta\,\Delta}}{2}\right)\left[\gamma\,
\bar{T}_{1} \,(2+A_{1}^{2})\right]\right.\nonumber\\&-&\left.
\left(\frac{Q_2^2\mathcal{Q}_{\Lambda\,\Lambda}+
2Q_1Q_2\,\mathcal{Q}_{\Delta\,\Lambda}
+Q_1^2\mathcal{Q}_{\Delta\,\Delta}}{2}\right)\left[\gamma\,
\bar{T}_{2} \,(2+A_{2}^{2})\right] \right\}  \nonumber
\\&\times&
\exp\left\{-\frac{P_1^2}{2}
\left[\frac{\bar{T}_{1}}{m}\,(2+A_{1}^{2})
\mathcal{R}_{\Lambda\,\Lambda}
+\frac{\bar{T}_{2}}{m}\,(2+A_{2}^{2})\mathcal{R}_{\Delta\,\Delta}
\right]\right\}\times \nonumber \\&\times&
\exp\left\{-\frac{P_2^2}{2}\left[\frac{\bar{T}_{1}}{m}\,(2+A_{1}^{2})
\mathcal{R}_{\Delta\,\Delta} +\frac{\bar{T}_{2}}{m}\,(2+A_{2}^{2})
\mathcal{R}_{\Lambda\,\Lambda}\right]\right\}\label{W}
\end{eqnarray}

The exact final result is given by:
\begin{eqnarray}
p^{ss}(x_1,v_1,x_2,v_2) & = &
\mathcal{G}_0\,\exp\left\{\mathcal{N}_{x_1x_1}x_1^2+\mathcal{N}_{x_2x_2}x_2^2
+\mathcal{N}_{v_1v_1}v_1^2+\mathcal{N}_{v_2v_2}v_2^2+
\right.\nonumber\\
&+& \left. \mathcal{N}_{x_1v_1}x_1v_1 +\mathcal{N}_{x_1x_2}x_1x_2
+\mathcal{N}_{x_1v_2}x_1v_2 +\mathcal{N}_{x_2v_1}x_2v_1
+\mathcal{N}_{x_2v_2}x_2v_2
+\mathcal{N}_{v_1v_2}v_1v_2\right\},\label{pss}
\end{eqnarray}
where all the coefficients $\mathcal{N}_{\alpha_{r}\beta_{s}}$ above
depend on the temperatures and mechanical constants of the system.

Due to the couplings present in Eq.(\ref{pss}), it does not describe
a usual Boltzmann distribution but instead a steady state where
couplings between position and moments arise. The basic reason for
this to occur is that the work done by the coupling spring is of the
form (work done on particle 1 by the spring) on an interval of time
$dt$
\[
dW_1\,=\,-k\,(x_1-x_2)\,v_1\,dt,
\]
generating, as we shall see, a correlation between $x_2$ and $v_1$ due to the coupling
above. A similar correlation between $x_1$ and $v_2$ also appears.

It is straightforward to show that these correlation functions are given by:
\begin{eqnarray}
\left<x_1v_2\right> &=& -\frac{\mathcal{D}_1}{\mathcal{D}_2},\label{x1v2}\\
\left<x_2v_1\right> &=&
-\frac{\mathcal{D}_3}{\mathcal{D}_2},\label{x2v1}
\end{eqnarray}
where
\begin{eqnarray}
\mathcal{D}_1 &=& 4\,\mathcal{N}_{x_1v_2}\, \mathcal{N}_{x_2x_2}\,
\mathcal{N}_{v_1v_1}-
\mathcal{N}_{x_1v_2}\,{\mathcal{N}_{x_2v_1}}^{2}-2\,
\mathcal{N}_{x_1x_2}\, \mathcal{N}_{x_2v_2}\,
\mathcal{N}_{v_1v_1}-\nonumber\\
&-& 2\, \mathcal{N}_{x_1v_1}\, \mathcal{N}_{x_2x_2}\,
\mathcal{N}_{v_1v_2}+ \mathcal{N}_{x_1v_1}\, \mathcal{N}_{x_2v_1}\,
\mathcal{N}_{x_2v_2}+ \mathcal{N}_{v_1v_2}\,\mathcal{N}_{x_1x_2}\,\mathcal{N}_{x_2v_1}\label{D1}\\
\mathcal{D}_2 &=& 16\, \mathcal{N}_{v_2v_2}\,
\mathcal{N}_{x_1x_1}\,\mathcal{N}_{x_2x_2}\,\mathcal{N}_{v_1v_1}-4\,{\mathcal{N}_{v_1v_2}}^{2}
\mathcal{N}_{x_1x_1}\, \mathcal{N}_{x_2x_2}-4\,
\mathcal{N}_{x_1x_1}\,{\mathcal{N}_{x_2v_2}}^{2}
\mathcal{N}_{v_1v_1}+4\,\mathcal{N}_{v_1v_2}\,\mathcal{N}_{x_1x_1}\,
\mathcal{N}_{x_2v_1}\,\mathcal{N}_{x_2v_2}-\nonumber\\
&-& 4\,
\mathcal{N}_{v_2v_2}\,\mathcal{N}_{x_1x_1}\,{\mathcal{N}_{x_2v_1}}^{2
}-4\,\mathcal{N}_{v_2v_2}\,\mathcal{N}_{x_2x_2}\,{\mathcal{N}_{x_1v_1}}^{2}-4\,{\mathcal{N}_{x_1v_2}}^{2}
\mathcal{N}_{x_2x_2}\,\mathcal{N}_{v_1v_1}+4\,\mathcal{N}_{v_1v_2}\,\mathcal{N}_{x_2x_2}\,\mathcal{N}_{x_1v_1}\,
\mathcal{N}_{x_1v_2}+\nonumber\\
&+&
{\mathcal{N}_{v_1v_2}}^{2}{\mathcal{N}_{x_1x_2}}^{2}+{\mathcal{N}_{x_2v_2}}^{2}{\mathcal{N}_{x_1v_1}}^{2}+4\,
\mathcal{N}_{x_2v_2}\,\mathcal{N}_{x_1x_2}\,\mathcal{N}_{x_1v_2}\,\mathcal{N}_{v_1v_1}-\nonumber\\
&-&
4\,\mathcal{N}_{v_2v_2}\,\mathcal{N}_{v_1v_1}\,{\mathcal{N}_{x_1x_2}}^{2}-2\,\mathcal{N}_{v_1v_2}\,
\mathcal{N}_{x_2v_2} \,\mathcal{N}_{x_1x_2}\,
\mathcal{N}_{x_1v_1}+{\mathcal{N}_{x_1v_2}}^{2}{\mathcal{N}_{x_2v_1}}^{2}-\nonumber\\
&-&
2\,\mathcal{N}_{v_1v_2}\,\mathcal{N}_{x_2v_1}\,\mathcal{N}_{x_1x_2}\,
\mathcal{N}_{x_1v_2}-2\,\mathcal{N}_{x_1v_1}\,\mathcal{N}_{x_1v_2}\,
\mathcal{N}_{x_2v_1}\,\mathcal{N}_{x_2v_2}+4\,\mathcal{N}_{v_2v_2}\,
\mathcal{N}_{x_2v_1}\,\mathcal{N}_{x_1x_2}\,\mathcal{N}_{x_1v_1},\label{D2}\\
\mathcal{D}_3 &=&4\,\mathcal{N}_{x_1x_1}\,\mathcal{N}_{x_2v_1}\,
\mathcal{N}_{v_2v_2}-\mathcal{N}_{x_2v_1}\,{\mathcal{N}_{x_1v_2}}^{2}-2\,
\mathcal{N}_{x_1x_2}\,\mathcal{N}_{x_1v_1}\,\mathcal{N}_{v_2v_2}-\nonumber\\
&-&
2\,\mathcal{N}_{x_1x_1}\,\mathcal{N}_{x_2v_2}\,\mathcal{N}_{v_1v_2}+\mathcal{N}_{x_2v_2}\,
\mathcal{N}_{x_1v_1}\,\mathcal{N}_{x_1v_2}+\mathcal{N}_{v_1v_2}\,\mathcal{N}_{x_1x_2}\,\mathcal{N}_{x_1v_2}\label{D3}.
\end{eqnarray}
Observe that we can swap $\mathcal{D}_1$ and $\mathcal{D}_3$ by
the transformation $1\,\leftrightarrow\,2$.

In  the equilibrium  limit $T_1=T_2=T$,  all the terms of the form
$\mathcal{N}_{xv}$ will vanish. In consequence, the couplings
represented in Eqs.(\ref{x1v2}) and (\ref{x2v1}) will also vanish.
In equilibrium the flux of heat ceases and velocities decouple from
positions, as in the following cases shown below.

\subsection{Interesting limits}
Two interesting limits arise. Firstly, by decoupling the particles
\[
k=0 \, \Rightarrow\,\mathcal{Q}_{\Delta\Delta}=
\mathcal{Q}_{\Lambda\Delta}= \mathcal{Q}_{\Delta\Delta} =
\mathcal{R}_{\Delta\Delta}=\mathcal{H}=0.
\]
The distribution is given by the product of two independent
Boltzmann terms:
\begin{equation}
\hspace{-3.0cm} p^{ss}(x_1,x_2,v_1,v_2)
=
{\frac {m\,k^{\prime}}{(2\,\pi)^2\,T_1T_2}} \,\,\exp\left\{-\frac
{k^{\prime}{x}^{2}_{1}}{2T_1} -{\frac{m{v}^{2}_1}{2T_1}}-\frac
{k^{\prime}{x}^{2}_{2}}{2T_2} -{\frac{m{v}^{2}_2}{2T_2}}\right\}.
\end{equation}

Secondly, by taking the equilibrium (same temperature) case
$T_1=T_2=T$. The final result corresponds to the Boltzmann
distribution:
\begin{equation}
\hspace{-3.0cm} p^{eq}(x_1,x_2,v_1,v_2)
=
{\frac {m\sqrt
{k^{\prime}(k^{\prime}+2k)}}{(2\,\pi\,T)^2}}\,\,\exp\left\{-\frac
{k^{\prime}x^{2}_{1}}{2T}-{\frac
{k^{\prime}{x}^{2}_{2}}{2T}}-\frac{k(x_{1}-x_{2})^2}{2T}
-{\frac{m{v}^{2}_1}{2T}} -{\frac{m{v}^{2}_2}{2T}}\right\}.
\end{equation}

%
%   VI
%
\section{Thermal conductance}
We are going to obtain the current of energy (heat) between the two
Brownian particles by two methods: the exact direct calculation of
the work rate between the particles,  and a Green-Kubo formalism
appropriate for finite systems~\cite{2009_JSTAT_L03001}.

The Green-Kubo (GK) formalism~\cite{livro_balescu}  has applications
for many problems such as fluid slab flow
properties~\cite{1994_PRE_49_3079}, diffusion in granular
fluids~\cite{2001_JStatPhys_105_723, 2002_PRE_65_051303,
2002_PRE_65_051304}, fluctuation-dissipation
theory~\cite{2008_JChemPhys_128_014504, 2008_PRE_77_051301}, thermal
conductance in condensed matter systems~\cite{2008_PRB_77_184302,
2008_JChemPhys_129_024507}, viscosity of trapped Bose
gas~\cite{2006_AoP_321_1063}, triple-point bulk and shear
viscosities~\cite{1980_PRA_22_1690,1988_PRA_38_6255} or
self-diffusion~\cite{1987_PRA_35_218} for Lennard-Jones fluids,
among others. The GK method depends crucially on the convergence of
time integrations of flux-flux correlation functions.

The convergence of the GK integral depends on the flux-flux
time-correlation functions decaying  fast enough, otherwise the time
integration will diverge such as happens for two dimensional
hydrodynamic systems~\cite{livro_resibois}. This is due to the
mode-coupling between hydrodynamic modes generating a
$t^{-1}$-dependent tail in the velocities correlation. However, for
three dimensions the tail goes as $t^{-3/2}$~\cite{livro_resibois,
1970_PhysRevA_1_18, 1970_JChemPhys_53_3813} and the Green-Kubo
integral converges. On the other hand, in one dimension
non-diffusive effects can affect the validity of Fourier's law while
a Green-Kubo approach might still be
valid~\cite{2003_PRE_67_015203}.

\subsection{Energy flux}

In order to proceed, we will  define the energies and fluxes for our
system. The ``local'' energy density will be given by
\begin{equation}\label{eq.54}
\epsilon_{1,2} = \frac12 m\,v_{1,2}^2+ \frac12 k^{\prime} x_{1,2}^2.
\end{equation}
The  contact of the particles with the thermal reservoirs, and the presence of the dissipative
terms, imply a flux of energy into, and out of, the system at both
positions. These instantaneous contact fluxes are given by~\cite{2008_AdvPhys_57_457}:
\begin{eqnarray}
j_{c1} & = & -\gamma v_{1}^2 + v_{1} \eta_1,\label{eq.55}\\
j_{c2} & = & -\gamma v_{2}^2 + v_{2} \eta_2.\label{eq.56}
\end{eqnarray}
As the coupling spring acts as the interaction channel between the
particles, we  define, for each particle, the transmitted heat
flux (Energy/Time) as
\begin{eqnarray}
j_{t1} & = & -k \left(x_1(t)-x_2(t)\right)v_{1},\label{eq.57}\\
j_{t2} & = & -k \left(x_2(t)-x_1(t)\right)v_{2}.\label{eq.58}
\end{eqnarray}
The local inter-particle elastic energy is defined as
\begin{equation}\label{eq.59a}
E_{el} =\frac12 k \left(x_1(t)-x_2(t)\right)^2.
\end{equation}
The total balance of energy requires that
the excess energy to be stored in the spring potential. Thus, it is
straightforward to see that the above definitions do respect energy balance since
\[
j_{t1} + j_{t2} = -dE_{el}/dt.
\]

The effective transfer flux $j_{12}$ can now be defined:
\begin{eqnarray}
j_{12} &=& \frac12 (j_{t1} - j_{t2})\nonumber \\
& = & -k \left(x_1(t)-x_2(t)\right)
\left(\frac{v_{1}(t)+v_{2}(t)}{2}\right).\label{eq.60}
\end{eqnarray}

The definition above corresponds to sharing the elastic energy,
defined in Eq.~\ref{eq.59a}, in equal parts between the
neighboring particles.

\subsection{Direct calculation of $\kappa$}

The thermal conductance is:
\begin{equation}\label{kappa}
\kappa\equiv\kappa(T,\Delta\,T) =
\frac{\partial}{\partial\,\Delta\,T} \langle
j_{12}\rangle_{\Delta\,T},
\end{equation}
where $A_1=A_2=0$, $T_1=T$, $T_2=T+\Delta\,T$, and
$\langle\,\rangle_{\Delta\,T}$ is the average at $\Delta\,T>0$. The
above expression for $\kappa$ goes beyond first order approximation
since it contains all the information needed to calculate the heat
flux, as shown in
\begin{equation}
\langle j_{12}\rangle_{\Delta\,T}\equiv \langle
j_{12}\rangle(T,\Delta\,T)= \int_{0}^{\Delta\,T}dt\,\kappa(T,t).
\end{equation}

The average heat flux is given by $\langle j_{12}
\rangle_{\Delta\,T}$ and can be calculated exactly:
\begin{eqnarray}
\left< j_{12} \right>_{\Delta\,T} &=&-k \left<\left(x_1-x_2\right)
\left(\frac{v_{1}+v_{2}}{2}\right)\right> =
-\frac{k}{2}\left(\left<x_1v_{2}\right>- \left<x_2v_{1}\right>
\right).
\end{eqnarray}
Using Eqs. (\ref{x1v2}) and (\ref{x2v1}), we write:
\begin{eqnarray}
\left< j_{12} \right>_{\Delta\,T}
&=&-\frac{k\,(\mathcal{D}_1-\mathcal{D}_3)}{2\,\mathcal{D}_2}
\end{eqnarray}
where the values of $(\mathcal{D}_1,\mathcal{D}_2,\mathcal{D}_3)$
are given in Eqs.(\ref{D1}), (\ref{D2}), and (\ref{D3}).

After some tedious (but straightforward) algebra, the final result
is rather simple:
\begin{equation}
\left< j_{12} \right>_{\Delta\,T} = 2 \,k\,\mathcal{H} \Delta  T
\Rightarrow\, \kappa = \frac{k^2\,\gamma\, }
{2\,\left[m\,k^{2}+\gamma^{2}(k+k^{\prime})\right]},\label{kappa1}
\end{equation}
where $\kappa$ is exact and independent of $T$ and $\Delta T$.

It is not unexpected to find the flux proportional to $\Delta\,T$,
since this result has been obtained for similar models
before~\cite{2004_JSP_116_783,2003_PhysRep_377_1,1970_PRA_4_1086}.
However, Eq.(\ref{kappa1}) represents the time-average over the full
dynamics of the system. We do not make any use of approximate
Master-equation-type methods, such as the Fokker-Planck
equation~\cite{livro_vankampen}, to obtain the value of $\kappa$.
Our method is equivalent to solving exactly the dynamical equations
of motion given the realization of the noise, then taking the noise
average, and finally time-averaging the final result.  In principle,
the present approach can be generalized for any type of noise, not
only white noise.  It is interesting to compare Eq.(\ref{kappa1})
with the results obtained from a Green-Kubo integration. This will
be an interesting test on the validity of the choice of the thermal
current, and also of the approximations used in order to derive the
Green-Kubo formalism.

\subsection{Green-Kubo calculation of $\kappa$}

The exact expression for $\kappa$ above can be compared with
proposals in the literature where Green-Kubo formulations for the
thermal conductance are given. In  the spirit of the previous
paragraph, the effective flux $\overline{j}$ plays the role of the
fluctuating flux $j_{12}$ for a Green-Kubo  relation
proposed~\cite{2009_JSTAT_L03001} for obtaining the thermal
conductance:
\begin{equation}\label{eq.63}
\kappa = \lim_{\Delta\,T\rightarrow0}
\frac{\langle\overline{j}\rangle_{\Delta\,T}}{\Delta\,T} =
\frac{1}{ \,T^2} \int_{0}^{\infty}dt\,
\langle\overline{j}(t)\overline{j}(0)\rangle,
\end{equation}
where  $\langle\,\rangle$ stands for the equilibrium average
($\Delta\,T=0$). We write
\begin{eqnarray}
\kappa &=& \lim_{\Omega\rightarrow\infty}\frac{1}{\Omega}
\int_{0}^{\Omega}dt\,\frac{1}{( T)^2} \int_{0}^{\infty}d\tau\,
\langle  j_{12}(t+\tau)j_{12}(t)\rangle_{\Delta T=0},\nonumber\\
&=& \lim_{z\rightarrow0^+}  \lim_{\theta\rightarrow 0^{+}}
\frac{z}{( T)^2}\int_{0}^{\infty}dt\,e^{-zt}
\int_{0}^{\infty}d\tau\, e^{-\theta\tau}\, \langle
j_{12}(t+\tau)j_{12}(t)\rangle_{\Delta T=0}.\label{eq.64}
\end{eqnarray}

Replacing the flux above into  Eq.(\ref{eq.64}), we obtain the
Green-Kubo expression for $\kappa$:
\begin{eqnarray}
\kappa &=&  \lim_{z\rightarrow 0^{+}} \lim_{\theta\rightarrow 0^{+}}
z\int_{0}^{\infty}dt\,e^{-zt}\,\frac{k^{ 2}}{T^2} \int_{0}^{\infty}
d\tau\, e^{-\theta\tau} \nonumber \times \nonumber
\\&\times& \left<\left[\left(x_1(t+\tau)-x_2(t+\tau)\right)
\left(\frac{v_{1}(t+\tau)+v_{2}(t+\tau)}{2}\right)\right]\right.\times \nonumber\\
&\times&\left. \left[\left(x_1(t)-x_2(t)\right)
\left(\frac{v_{1}(t)+v_{2}(t)}{2}\right)\right] \right>_{\Delta
T=0}\label{eq.59}
\end{eqnarray}

\begin{figure}[tbh]
\begin{center}
\includegraphics[width=0.5\columnwidth,angle=0]{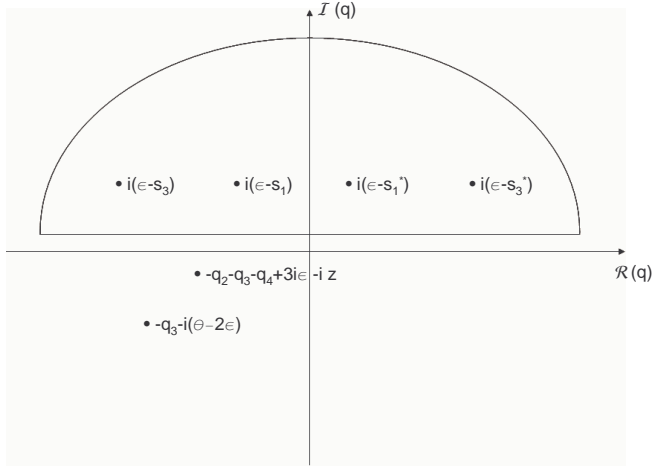}
\end{center}
\caption{Integration path over the poles for the Green$-$Kubo
calculation of the conductance.}\label{fig2}
\end{figure}

After some algebraic manipulation the expression for the thermal
conductance becomes:
\begin{eqnarray}
\kappa &=&
\lim_{z\rightarrow 0^{+}} \lim_{\theta\rightarrow
0^{+}}\lim_{\epsilon\rightarrow 0^{+}} \frac{k^{ 2}}{16\,T^2}
\int_{-\infty}^{\infty}\frac{dq_1}{2\pi}
\int_{-\infty}^{\infty}\frac{dq_2}{2\pi}
\int_{-\infty}^{\infty}\frac{dq_3}{2\pi}
\int_{-\infty}^{\infty}\frac{dq_4}{2\pi}\times \nonumber \\&\times&
\frac{z}{z-\left[ (iq_1+\epsilon)+ (iq_2+\epsilon)+ (iq_3+\epsilon)
+(iq_4+\epsilon)\right]}\,\,\,
\frac{(iq_3+\epsilon)(iq_4+\epsilon)}{\theta-\left[ (iq_1+\epsilon)
+
(iq_3+\epsilon)\right]}\times\nonumber\\
&\times&
\left<\left[\left(\tilde{x}_1(iq_1+\epsilon)-\tilde{x}_2(iq_1+\epsilon)\right)
\left(\tilde{x}_{1}(iq_3+\epsilon)+\tilde{x}_{2}(iq_3 +\epsilon)\right)\right]\right.\times \nonumber\\
&\times&\left.
\left[\left(\tilde{x}_1(iq_2+\epsilon)-\tilde{x}_2(iq_2+\epsilon)\right)
\left(\tilde{x}_{1}(iq_4+\epsilon)+\tilde{x}_{2}(iq_4+\epsilon)\right)\right]
\right>\nonumber\\
&=& \lim_{\theta\rightarrow 0^{+}}\lim_{\epsilon\rightarrow 0^{+}}
\frac{k^{ 2}\gamma^2}{4}
\int_{-\infty}^{\infty}\frac{dq_1}{2\pi}\int_{-\infty}^{\infty}\frac{dq_3}{2\pi}
\frac{(iq_3+\epsilon)(-iq_3-\epsilon)}{\theta-\left[ (iq_1+\epsilon)
+
(iq_3+\epsilon)\right]}\times\nonumber\\
&\times&
\frac{1}{[\Gamma(iq_1+\epsilon)+k][\Gamma(-iq_1-\epsilon)+k][\Gamma(iq_3+\epsilon)-k][\Gamma(-iq_3-\epsilon)-k]}.
\label{eq.D3}
\end{eqnarray}
where the poles are given by:
\[
s_{1}=-\frac{\gamma}{2m}+\,i
\sqrt{\frac{k^{\prime}}{m}-\frac{\gamma^2}{4m^2}}; \,\,\,\,s_{3}=
-\frac{\gamma}{2m}+\,i
\sqrt{\frac{(2k+k^{\prime})}{m}-\frac{\gamma^2}{4m^2}},
\]
and the integration path is shown in Fig.\ref{fig2}.

Equation (\ref{eq.D3}) gives exactly the same result of
Eq.(\ref{kappa1}), showing that both approaches are completely
consistent.

%%%%%%%%%%%%%%%%%%%%%%%%%%%%%%%%%%%%%%%%%%%%%%%%%%%
%%%%%%%%%%%%%%%%%%%%%%%%%%%%%%%%%%%%%%%%%%%%%%%%%%%

\subsection{Discussion}

The coherence shown  for the  thermal conductance results for a
finite systems, calculated either directly, Eq.(\ref{kappa1}), or
via the Green-Kubo approach, Eq.(\ref{eq.64}), seems to point to the
validity of considering the microscopic work as the correct
fluctuating flux variable to be used for coupled particle systems.
In fact, for more realistic models in which the number of particles
is large, solving the same problem  for non-harmonic potentials
might be the way to obtain a rigorous demonstration of Fourier's
Law.

In our case, despite the somewhat involved aspects of the algebra,
the final value for $\kappa$ is quite simple and carries the
influence of both couplings, $k$ and $k^{\prime}$, the friction
coefficient $\gamma$, and the inertia $m$. The program we followed
in order to find $\kappa$ is equivalent to solving the exact
equations of motion of the Brownian particles system for each
realization of the noise functions, and then taking the average over
the noise.  No approximations of any sort are necessary once the
basic model is provided. The present treatment can be extended to
other distinct kinds of noise, such as colored noise
(non-Markovian), or even distinct heath baths acting on the same
particles.

However, the present method can readily be  extended to (finite)
systems composed of more than two Brownian particles, systems that
may be large enough to be taken as ``macroscopic''. The difficulties
to treat such systems are operational or numerical, rather than
conceptual.

\section{Conclusions}

Brownian particles (BP) are an excellent laboratory for studying
non-equilibrium physics. They are simple to describe but present
many of the features of more complex models, such as the possibility
of reaching stationary states when submitted to thermal contacts  at
distinct temperature. They are also good approximations for larger
systems, like polymers, that could be modeled by chains of BP
attached to each other by some type of attractive potential.

Another interesting characteristic of such systems is that they
are simple enough so that we can extract exact solutions for their
long-time behavior. This allows us to obtain results that are hard
to come by using other methods. It is already known that we can
obtain exactly the equilibrium probability distribution for
Brownian particles subjected to Markovian or non-Markovian noise,
or a combination of both. This type of external forcing allows us
to keep a system formed by a single particle constantly on an out
of equilibrium steady-state.

Furthermore, techniques based on time-averaging are very
interesting since they are ensemble independent, driven only by
the dynamical relations governing the interaction Brownian
particle-heath bath. In fact, this corresponds to following a
system during the realization of an experiment.

In the present work, we have studied the thermal  conductance for a
system of coupled particles, by taking advantage of the mechanically
simple characteristics of Brownian particles and of time-averaging.
Our system consists of two particles coupled by a spring potential,
with the heat flux flow $j_{12}$ being due to the mechanical work
done through the spring coupling the two particles. The particles
may be kept in contact with distinct thermal baths and the flow of
energy may be obtained by means of the thermal conductance
coefficient $\kappa$, appropriate for small systems. The latter is
calculated both from first principles, and by means of a Green-Kubo
formulation,  and the obtained exact results are identical. The
final form of $\kappa$ depends only on the variables of the system,
such as the mass of the particles, the spring couplings and the
friction coefficients. The thermal conductance can be thought as a
first step for obtaining the equivalent form for, more
sophisticated, macroscopic systems such as long polymers.

We believe that the coherence between the exact direct  calculation
and the exact Green-Kubo formulation for $\kappa$ shows the
correctness of the basic definitions, in special that  of the heat
flux, used in the problem.

\acknowledgments One of us W.A.M.M. thanks the Brazilian funding
agencies Faperj and CNPq and D.O.S.P. would like to thank the
Brazilian funding agency CNPq for the financial support.

%\newpage

%\section{References}

\end{document}